# Managers' Choice of Disclosure Complexity

Jeremy Bertomeu,[*] *Washington University in St Louis*

**Abstract**

Aghamolla and Smith (2023) make a significant contribution to enhancing our understanding of how managers choose financial reporting complexity. I outline the key assumptions and implications of the theory, and discuss two empirical implications: (1) a U-shaped relationship between complexity and returns, and (2) a negative association between complexity and investor sophistication. However, the robust equilibrium also implies a counterfactual positive market response to complexity. I develop a simplified approach in which simple disclosures indicate positive surprises, and show that this implies greater investor skepticism toward complexity and a positive association between investor sophistication and complexity. More work is needed to understand complexity as an interaction of reporting and economic transactions, rather than solely as a reporting phenomenon.

Aghamolla and Smith (2023) offer a valuable fresh perspective on the management of complexity in financial reporting. By taking on the challenge of clarifying a difficult topic for a general audience, it is a perfect example of what theory is best suited for: provide insight about the interpretation of complexity as a choice. The model is an extension of Dye (1985) with multiple messages. A *simple* message is understood equally by all investors, whereas a *complex* message may be designed to obfuscate or inform. The authors assume a ranking between the sensitivity of prices for each message, with the complex obfuscated (informative) message affecting prices the least (most) and the simple message having an intermediate sensitivity. The firm increases price sensitivity as the news becomes more favorable. This translates into a strategic complexity equilibrium in which simple disclosures are chosen for moderate news, while complex obfuscated (informative) messages are chosen for bad (good) news.

Such endogenous complexity presents an intriguing consideration that has implications about how financial reporting may prey on investors unable to adequately process information. Indeed, accounting is a self-evident application of complexity, drawing from the simplest building blocks from debits and credits to recreate the entire structure of a nexus of contracts (Jensen 1983). This brings us to a key point emphasized in the study. How do firms strike the right balance between providing valuable information to investors and avoiding unnecessary complexity or the perception of intentional deception?

[*]Contact Author. Jeremy Bertomeu, Associate Professor, Olin Business School, Washington University in St Louis, One Brookings Drive St. Louis, MO 63130-4899, ph. 314-935-6000, email: bjeremy@wustl.edu. I gratefully thank the editor (John Core), the authors and conference participants for many suggestions and comments that have greatly benefited the discussion.

To address this question in several steps, my discussion is organized in three main sections. Section 1 explains the intuitions of the model with simple graphical arguments, recovering the necessary assumptions, economic tradeoffs and equilibrium characterization with minimal mathematical formalism. Section 2 provides a qualified critique of how these insights are weaved together, noting that five parameters are put to work to obtain one main insight. I then formulate the model in terms of a single parameter capturing investor savviness toward complexity and offer two predictions: (i) investors discount complex reports, and (ii) investor sophistication and complexity are positively associated. Section 3 revisits the empirical implications of the model, and further clarifies several testable hypotheses that may better organize and motivate empirical work using proxies of complexity.

# 1 The Aghamolla and Smith Framework

I present below the assumptions underlying the Aghamolla and Smith (2023) approach to strategic complexity, and how its primary intuition can be illustrated from simple graphical arguments. The model is based on the voluntary disclosure paradigm of Verrecchia (1983) and Dye (1985). This paradigm can be recognized as a class of sender-receiver games with two building blocks:

(1) *Belief maximization*. The manager is solely interested in maximizing the stock price (the perception of the manager's private information by competitive investors);

(2) *Type independent cost*. A disclosure is a choice over a set of messages whose feasibility, but not their cost, is a function of the manager's private information.

Assumption (1) excludes any agency considerations in which a manager may be involved in other decisions affecting firm value and further rules out that owners may use other means to align actions toward long-term shareholder value. Importantly, the firm cannot set a long-term contract with its management that commits it to be transparent. Assumption (2) is more subtle and sometimes leads to the approach being, reductively, placed under the umbrella of truthful disclosure. Rather, the manager sends messages that depend on their information, but they do not have the ability to send any message, as in cheap talk, or spend costly effort to send messages that they ought not send, as in earnings management models.[1] Cost independence includes sending the actual message (a verifiable fact) but can also feature

---

[1]It is possible in these models that some managers make either their "truthful" disclosures, or some another disclosure that mimics better information, see, for example, Sansing (1992) or Marinovic (2013). While this literature was initially

vague messages or non-disclosure (i.e., a message available to all regardless of their information).

To see how the model fits into the paradigm, consider the starting point of the Dye (1985) model, in which the manager may disclose, but some managers are constrained to stay silent, either because they either do not know the information or do not have the means to convey it credibly. The resulting reporting space is binary. One message (non-disclosure) implies that prices do not depend on the private information. The other message (disclosure) implies a positive sensitivity of prices to the private information. The price-maximizing strategy is given by the maximum of these two choices, which readily leads to a threshold equilibrium above (below) which the manager prefers to disclose (stay silent).

In the current model, managers do not choose between non-disclosure and disclosure, but there are still two messages causing prices to be differentially sensitive to the private information. The first message "complex uninformative" (obfuscation when chosen in lieu of an informative disclosure) takes the place of non-disclosure and implies a low price sensitivity because only a subset of sophisticated investors respond to the message and may fail to read through the complexity. The second message "complex informative" replaces disclosure and yields a higher sensitivity than its uninformative counterpart because sophisticated investors can always recover the information.

To avoid an unravelling equilibrium in which all managers choose their more informative option, AS assume that with some probability the manager is constrained to disclose in a certain way. The solution to this problem involves picking the maximum of a piecewise linear surface, with complex uninformative disclosures for low private information and complex informative ones for high private information. The resulting equilibrium is plotted in the left-hand side of Figure 1 and demonstrates that the main intuition for a standard threshold equilibrium holds: it is the *relative* slope of the two messages that matters, rather than, as imposed in the Dye (1985) model, the price response being flat conditional on the least informative message.

A third message is then introduced, which is later labelled as "simple" and features a sensitivity to information that is somewhere in-between that of complex informative and uninformative. A simple disclosure provides the manager's private information with some probability and is uninformative otherwise. All investors (including those that are sophisticated) get the information from the simple disclosure if it is informative. To fine tune the model to this happy spot of in-between sensitivity, the probability that the simple disclosure is informative must be lower than the fraction of sophisticated investors but greater than the chance that sophisticated investors understand a complex uninformative disclosure. I draw a diagram in the right hand-side of Figure 1 (corresponding to Figure 2 in the study) the complete strategic

---

developed in accounting, it has seen a recent resurgence in finance (Frenkel, Guttman and Kremer 2020, Aghamolla and An 2021, Banerjee, Marinovic and Smith 2022, Kremer, Schreiber and Skrzypacz 2022).

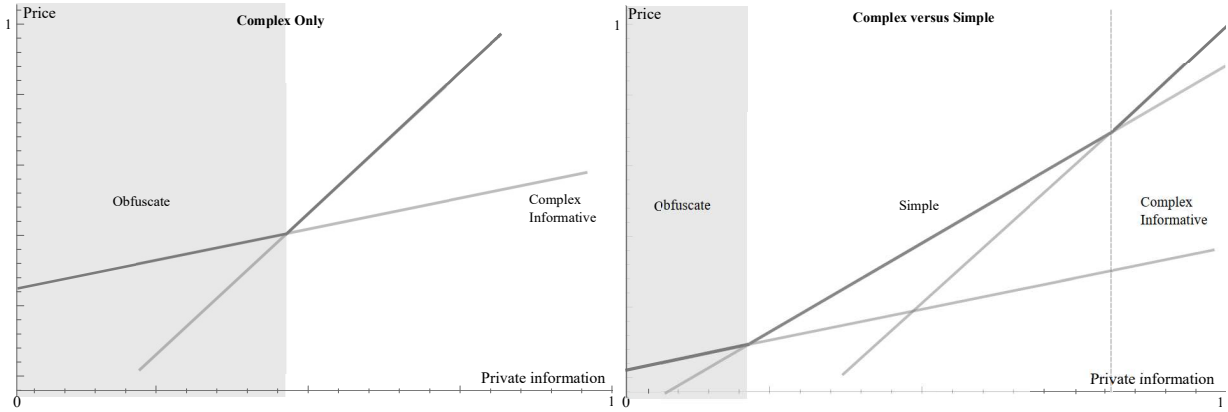

Figure 1: Complex only (left) versus strategic complexity equilibrium (right)

complexity equilibrium by adding to the plot the market price conditional on a simple disclosure. The maximum of these lines traces a third intermediate region of simple disclosures, that are neither bad enough to trigger obfuscation, nor good enough to call for an informative disclosure.

Although all equilibria take this form, the model may have multiple equilibria that differ according to whether investors perceive complex disclosures (on average) as good or bad news. For background, communication games in which a partially-informative message does not fully reveal the information have multiple equilibria (Stocken 2013, Hart, Kremer and Perry 2017). Here, unsophisticated investors see only the presence of a complex disclosure but do not know the underlying information. Therefore, even after a disclosure, they form a price based on whom they conjecture will choose that type of disclosure. Thus, if they expect more obfuscators to choose complex disclosure, they may become skeptical of complex disclosures, further deterring managers with better news from choosing complex disclosures and implying an equilibrium in which complex disclosures are bad news. The opposite may occur if managers with worse news are expected to prefer simple disclosures, implying simple disclosures being bad news.

The price sensitivity to complex informative disclosures is greater than all other disclosures, so a conceivable equilibrium conjecture is that complex disclosures are good news because they tend to made for better news. Following this intuition, there is always an equilibrium in which complex disclosures are perceived to be better than simple disclosures. There is an economic force in the model that tends to balance overly skeptical expectations against the potentially more informative complex disclosure. I refer to this as a "simple bad news" equilibrium and it implies that investors view simple disclosures as a form of strategic obfuscation, as illustrated in Figure 2.

If, however, we weaken this force so that simple disclosures have nearly the same price sensitivity as complex informative disclosures, there is another self-fulfilling equilibrium in which managers with bad news flock to obfuscating complex disclosures, making complex disclosures bad news. In summary, when simple disclosures are sufficiently informative, there is a second equilibrium in which simple disclosures are more widely used and more favorably received than complex disclosures, i.e., "simple good news." In this equilibrium, shown shaded in Figure 2, simple disclosure payoffs are shifted upward while complex disclosure payoffs are shifted downward, implying that simple disclosures are made more often and for more positive news.

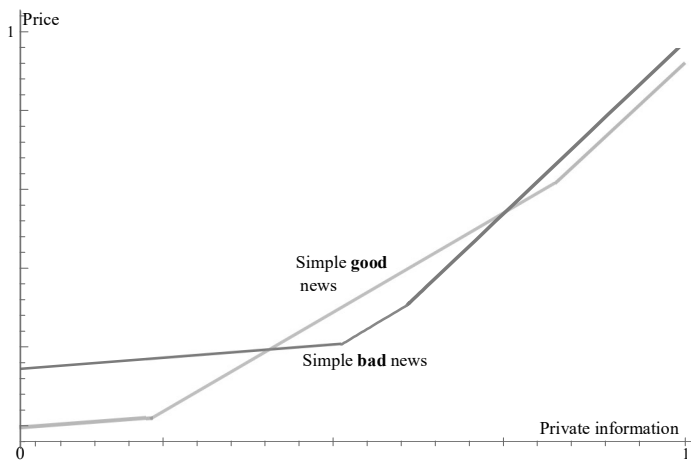

Figure 2: Multiple strategic complexity equilibria

## 2 A Simpler Model

Environments featuring complexity are inherently difficult to study, and the model offers an attempt at a descriptively richer depiction of various moving parts. However, for purposes of parsimony and to lay out the intuition as succintly as possible, this deliberate modelling choice may not make clear the minimal components that produce the most important predictions.

To illustrate this point, note that the model requires a minimum of five suitably fine-tuned parameters: the probabilities of being exogenously compelled to be simple ($\beta\omega_S$) or complex uninformative ($\beta\omega_U$), the fraction of sophisticated investors ($\chi$), and the probabilities of understanding simple ($\rho_S$) or complex disclosures ($\rho_U$). Table 1 below summarizes the probabilities of observing the firm's disclosure $y$ by investor sophistication and expected price.

The model further assumes supplementary structure on how messages can be processed, requiring that simple disclosures cannot be obfuscated and that unsophisticated investors are able to correctly separate complex versus simple disclosures. Finally, the solution requires relatively complex higher-order beliefs, even on the part of unsophisticated investors, not all of whom may be able to fully understand or play the rational expectations game (Corgnet, Desantis and Porter 2018, Corgnet, Deck, DeSantis and Porter 2022). Investors must collectively form three non-disclosure beliefs to price the firm following (i) a simple but uninformative event , (ii) a complex disclosure perceived by an unsophisticated investor, and (iii) an obfuscated complex disclosure. In comparison, in standard disclosure theory, there is only a single friction parameter and a non-disclosure belief.

| **Disclosure** | **Soph.** | **Unsoph.** | **Expected Price = χSoph + (1-χ)Unsoph** |
|---|---|---|---|
| Simple | $\rho_S$ | $\rho_S$ | $\rho_S y + (1-\rho_S)\mathbb{E}(\tilde{y}|x=S)$ |
| C-Informative | 1 | 0 | $\chi y + (1-\chi)\mathbb{E}(\tilde{y}|x=C)$ |
| C-Obfuscated | $\rho_U$ | 0 | $\chi\rho_U y + (\chi(1-\rho_U)\mathbb{E}(\tilde{y}|x=O) + (1-\chi)\mathbb{E}(\tilde{y}|x=C))$ |

Table 1: Notation, with S (simple), C (complex) and O (obfuscated)

In this section, I simplify the model, starting from a structure that reduces the number of parameters to just one to better understand the model mechanics. Let $\tilde{y}$ denote the manager's information about the firm value (earnings) and, for the sake of exposition, assume that it is uniformly distributed on $[0,1]$. As in the baseline model, the manager has three options:

(a) a simple disclosure yields a binary signal "good" if $y$ is above average (i.e., a positive earnings surprise), and "bad" otherwise;

(b) a complex informative disclosure reveals $y$ with probability $q \in (0,1)$ and, otherwise, yields no information, implying an expected price[2]

$$P_{CI}(y) \equiv qy + (1-q)P_\emptyset, \quad (2.1)$$

where $P_\emptyset$ is the price conditional on non-disclosure;

(c) a complex obfuscated message yields no information with probability one, implying a price $P_\emptyset$.

Above, I reduce the complexity of Aghamolla and Smith (2023) in several important ways. First, I assume that a simple disclosure is well-understood but imprecise; this is modelled here in terms of coarseness, whereas the original model assumes that the message may be lost with positive probability. My formulation simplifies the inference process because (i) no manager would voluntarily send the bad simple message, and (ii) the set of managers choosing to send the good message should be anticipated to be in the form of an interval $[1/2, \tau)$, where $\tau$ is the threshold above which the manager switches from a simple good message to a complex informative disclosure. Taking expectations, the conjectured price conditional on an equilibrium simple message is

$$P_S \equiv \frac{1/2 + \tau}{2}. \tag{2.2}$$

---
[2] I assume that complex disclosures are used for above-average events. Similar results can be obtained if we assume that the message can be sent for both good and bad news. In this case, there may be some additional complex informative disclosures made below the threshold $1/2$.

This price depends only on the marginal manager $y = \tau$ indifferent to a complex informative disclosure; therefore, evaluating at $P_C(\tau)$, this indifference condition must further imply

$$P_\emptyset \leq \underbrace{\frac{1/2 + \tau}{2}}_{=P_S} = \underbrace{q\tau + (1-q)P_\emptyset}_{=P_{CI}(\tau)}. \tag{2.3}$$

Note that there are no investor types in this model, whereas in the original model some investors may better understand complex informative disclosures. However, Aghamolla and Smith also assume that the price averages out the belief of each investor, so their resulting pricing equation is very similar to my equation (2.1); $1 - q$ can be interpreted as the probability of failing to process the information from the standpoint of a representative investor or the weight assigned to a mass of unsophisticated investors as in their model. Following this interpretation, the parameter $q$ may be used as a proxy for the overall degree of investor sophistication.

The parameter $q$ serves a dual function as a friction that prevents unravelling, since complex informative disclosures may fail and pass as uninformative. Therefore, without adding any extra parameter, we can solve the model using Bayes rule to derive the price conditional on no-disclosure:

$$P_\emptyset = \frac{1/2 \times \mathbb{E}(\tilde{y}|\tilde{y} \leq 1/2) + (1-\tau)(1-q)\mathbb{E}(\tilde{y}|\tilde{y} \geq \tau)}{1/2 + (1-\tau)(1-q)}. \tag{2.4}$$

To decompose the conditional expectation, investors know that non-disclosure could be due to strategic obfuscation, which occurs when a manager observes a value below $1/2$, or to unreadable but genuine complex disclosure, which occurs with a probability $(1-\tau)(1-q)$ that the value is above $\tau$ but the disclosure is uninformative. Solving for $P_\emptyset$ from (2.3) and substituting in (2.4) yields a quadratic form:

$$1 - 2\tau^* + \frac{4q(1-q)}{q(7-4q) - 2}(\tau^*)^2 = 0, \tag{2.5}$$

where, for a valid solution to exist (so that the simple disclosure region $[1/2, \tau]$ is not empty), it must

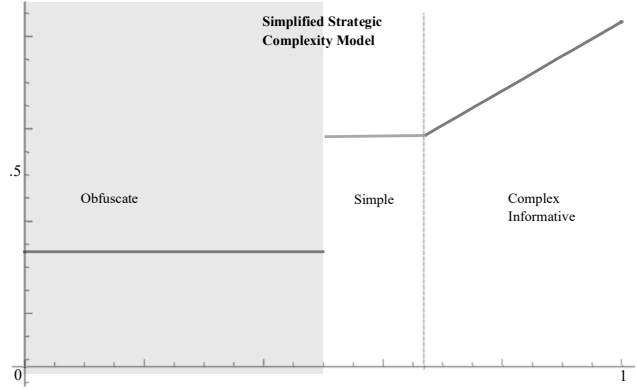

Figure 3: Simplified Analysis (with $q = 3/4$)

hold that $q > 2/3$. Hence, the model has a unique solution given by

$$\tau^* = \frac{1}{1 + \sqrt{\frac{3q-2}{q(7-4q)-2}}}. \tag{2.6}$$

In Figure 3, the strategic complexity equilibrium in this simplified model unveils several differences compared to the baseline model. Most importantly, it is unique and exhibits a distinct characteristic: A simple disclosure is unambiguously interpreted as good news, whereas in the robust equilibrium of the model, simple disclosures are perceived as bad news. Moreover, the model yields a clear and intuitive prediction: in equation (2.5), the probability of a simple disclosure decreases as investor sophistication increases. The reasoning underlying this prediction is straightforward, as a more sophisticated investor base tends to value and prefer the more informative message over a simple one.

## 3 Empirical Implications

I discuss next the empirical hypotheses generated by the theory and present the three stated empirical implications, as well as their inherent limitations and potential threats to testability. A strategic complexity equilibrium implies that complexity is nonmonotonic in the private information known to the manager. The manager should make simple disclosures for moderate news and complex disclosures for extreme news, either favorable (informative) or unfavorable (obfuscating). This intuition likely holds more generally over more messages and can be expressed as follows: The manager chooses a message whose price sensitivity increases in the reported news, putting more emphasis on affecting prices for more favorable news. I summarize this prediction in terms of the following hypothesis based on Proposition 1.

**H1.** The relationship between complexity and news is U-shaped, with greater observed complexity for both sufficiently positive and sufficiently negative news, and simple disclosures for moderate news.

As noted earlier, H1 holds only if the informativeness of simple disclosures is higher than the informativeness of complex uninformative disclosures and lower than that of complex uninformative disclosures. If simple disclosures become more (less) informative, simple disclosures will occur over better (worse) news and complex disclosures will occur over worse (better) news. These equilibria place simple disclosures for news either above or below all types of complex disclosures, implying a monotonic relation between disclosure and news. In this case, there is no U-shape.

The value of a theory is not only in determining whether it is true or false, but also in whether the theory provides broader insights into other questions and research designs. If H1 is true, this has critical implications for studies that use complexity as a control variable. Because complexity endogenously captures private information about the firm's future prospects, controlling for complexity may eliminate the source of variation necessary to test a theory in which decisions are influenced by information asymmetry or future prospects. As an example, consider a decision to manipulate earnings near a performance measure where both complexity and the extent of manipulation are a function of fundamentals. In this example, controlling for complexity can eliminate the strategic component in a manipulation proxy and amplify an endogeneity problem because controls are correlated with outcomes (Bertomeu, Beyer and Taylor 2016). To emphasize this point, controlling for complexity in AS is similar to controlling for forward-looking private information, such as a management forecast.[3] The framework provides a theoretical rationale that helps address the use of complexity in research studies. Controls for complexity should not be automatically invoked as a solution or treated as a panacea. When complexity is included as a control variable in a setting where private information is part of the hypothesized causal mechanism, it is essential to ensure that the findings derived from the study remain robust even when complexity is excluded as a control.

A second implication of the theory is that more sophistication reduces the propensity to issue complex disclosures. This is puzzling because, after all, sophisticated investors increase the quality of the informative disclosures and therefore *managers should gravitate to the more informative communica-*

---

[3] As another example, according to H1, complexity may also indicate more uncertainty (since it is chosen in response to more extreme events). As a result, such a control would be inappropriate in studies seeking a causal statement about uncertainty. Unfortunately, it has become a common practice in empirical research to justify control variables on the basis that a previous empirical study used the same controls. This ongoing practice is highly problematic because the use of controls without justification turns the process of conducting multivariate regressions into a set of conventions such that no good argument can be given to prove or disprove that a control is justified. A control should be justified by reference to a theory, evidence that the theory is empirically supported, and a possible argument that the identifying variation in that alternative theory is separate

*tion*. Note further that the simplified model of Section 2, with some moving parts of the model removed, predicts that greater investor sophistication decreases the probability of simple disclosures, consistent with intuition but not the result presented in AS. A key driver of this result seems to be that managers indifferent between complex uninformative and simple disclosures shift toward simple disclosures, because simple disclosures are now relatively more effective at hiding bad news. This in turn may increase propensity to issue simple disclosures. I state this hypothesis from Figure 5 in Aghamolla and Smith (2023) below.

**H2.** Managers are less likely to choose complex disclosures, when investors are more sophisticated or simple disclosures are less informative.

Proving the formal comparative static is complicated and, while numerical examples with the normal or uniform distribution generally support that H2 is true, there are countervailing forces that make the intuition less obvious. Recall that from the perspective of unsophisticated investors, complex disclosures are perceived as a mixture of very favorable news and obfuscation: The higher the sophistication, the more obfuscators tend to opt for simple disclosures for fear of being discounted by sophisticated investors. This is another reason why less complex information is disclosed after an increase in sophistication, but it also has an opposite effect. Since investors now expect less obfuscation, beliefs may be more favorable after complex disclosure, which in turn encourages more favorable managers to choose complex informative disclosures. This counter-effect appears to be second-order relative to the other effects, in that it is a consequence of the comparative static presented in H2, and I believe this is why it is dominated for many distributions.[4]

But there is also the question of whether the hypothesis is predicted by the theory if we do not hold constant which equilibrium of the game is played when investor sophistication changes. Recall that the game can have two equilibria: a simple good news and a simple bad news equilibrium. Empirical evidence generally supports the simple good news equilibrium, in which investors tend to extract less information, exhibit reduced willingness to trade, and apply a higher discount rate to complex messages (Li 2008, Blankespoor, deHaan and Marinovic 2020). In the model, however, this equilibrium exists only in a narrow band of parameter values such that investor sophistication is low (so that simple disclosures achieve a price sensitivity sufficiently similar to complex informative disclosures). As investors become more sophisticated, the simple good news equilibrium will cease to exist because the complex informative message becomes too powerful at separating managers with better news. At this point, managers shift to the simple bad news equilibrium implying a lower propensity to adopt simple disclosures, contradicting H2. The hypothesis must then be formulated with one supplementary condition: assuming that sophistication is sufficiently low (so that the simple good news equilibrium exists)...

---

[4] So, if not a complete proof, it may be a conjecture for which, as French mathematician Pierre de Fermat puts it, one has "a truly marvelous demonstration of this proposition which this margin is too narrow to contain."

... but not too low. If investors sophistication becomes so low that markets are more responsive to simple disclosures, the prediction will switch to a strategic simplicity equilibrium in which simple disclosures are used to convey good news. This equilibrium implies that simple disclosures become even more desirable, leading to greater use of simple disclosures when investor sophistication is lower. Unless we measure the price sensitivity to different types of news, assessing H2 is partly an empirical question within the scope of the framework. Note further that the simplified model of Section 2 predicts that more investor sophistication *decreases* the probability of simple disclosures.

The last important implication of the model is that, in the robust equilibrium, simple disclosures are bad news, which I explain heuristically below. Recall that the best hope for a manager with bad news is to pretend to be average by pooling with constrained managers with no reporting discretion. Such pooling, however, is unlikely to benefit above-average managers, who should then choose their most informative "complex-informative" disclosure when they are able to do so. This leads to the conclusion that there is always a simple bad news equilibrium, in which only managers with below-average news would make simple disclosures. Thus, regardless of sophistication, there is an equilibrium in which markets react negatively to *all* simple disclosures.[5] The following statement corresponds to Corollary 1 and Figure 4 in Aghamolla and Smith (2023).

**H3.** Simple disclosures trigger a negative price reaction and tend to indicate future unfavorable events.

I will argue that H3 is an important insight, but it should be viewed less as a falsification test and more as part of a full test of the model. Empirically, complex disclosures tend to be perceived negatively by investors on average (Li 2008, Lo, Ramos and Rogo 2017, Blankespoor et al. 2020). While there may be more empirical analyses to be made on this point, the evidence thus suggests that the simple good news equilibrium is the empirically relevant equilibrium and, moreover, is consistent with the prediction of the simplified model in Section 2 that simple disclosures are, on average, better news than complex disclosures. Further, the model does not strongly point to the simple bad news equilibrium as robust, as this equilibrium will cease to exist for other parameter values and there may be parameter values outside of the range examined in which there are only equilibria such that the market responds negatively to complexity.

---

[5]As in AS, I use the term of "robust" because this equlibrium exists for all parameter values examined in the study.

# 4 Concluding Remarks

Aghamolla and Smith develop a critical contribution to our understanding of strategic complexity in financial reporting. What we know about complexity has been driven primarily by empirical facts, while the development of theories of complexity has lagged behind. In this discussion, I lay out the main assumptions and implications of the theory, present a simplified tractable version of the model, and explain the primitive conditions under which the theory implies an association between price and complexity, a U-shape relationship between complexity and returns, and a negative association between complexity and investor sophistication.

While the approach holds promise for future empirical investigation, complexity may exhibit characteristics that currently remain outside the framework. In particular, the assumptions draw a strict distinction between reporting and transactional complexity, by considering the underlying economics and decision-making processes as fixed. This conceptual distinction can be beneficial for theoretical purposes, but in practical terms, may prove artificial, as managers constantly innovate and encounter new transactions that inherently introduce new complexity (Bertomeu and Magee 2011). Indeed, many empirical proxies of complexity, such as the number of divisions or foreign operations, XBRL tags, or accounting treatments (Blankespoor et al. 2020) jointly capture both transactional and reporting complexity in ways that cannot be easily disentangled. Incorporating the economic role of complexity may further extend the theory to understand the relationship between real effects or decision making and complexity, and how complexity helps or hinders the resolution of agency problems via long-term contracts or other mechanisms. I conclude that more work is needed to address these gaps in our understanding of the concept.